\documentclass[aps,prb,reprint,superscriptaddress,flatfix]{revtex4-2}
\usepackage{comment}
\usepackage{bm}
\usepackage[final]{graphicx}
\usepackage{epsfig}
\usepackage{color}
\usepackage{verbatim}
\usepackage[OT1]{fontenc}
\usepackage{dcolumn}
\usepackage{xcolor}
\usepackage[final]{graphicx}
\usepackage{ragged2e}
\usepackage{amsmath}
\usepackage{bm}
\usepackage{floatrow}
\usepackage{physics}
\usepackage{times}
\usepackage[colorlinks=true,allcolors=blue]{hyperref}
\usepackage{relsize}
\usepackage[capitalise]{cleveref}
\usepackage{xcolor}
\usepackage{fixltx2e}
\usepackage{xspace}
\usepackage{multirow}

\usepackage{adjustbox}
\usepackage{booktabs}
\usepackage{makecell}
\usepackage[T1]{fontenc}

\usepackage{xr}
\makeatletter
\newcommand*{\addFileDependency}[1]{%
  \typeout{(#1)}
  \@addtofilelist{#1}
  \IfFileExists{#1}{}{\typeout{No file #1.}}
}
\makeatother

\newcommand*{\myexternaldocument}[2][]{%
    \externaldocument[#1]{#2}%
    \addFileDependency{#2.tex}%
    \addFileDependency{#2.aux}%
}

\myexternaldocument[S-]{supp}

\newcommand{\LMetalO}{Li\textsubscript{2}MO\textsubscript{3}\xspace}
\newcommand{\LTMO}{Li\textsubscript{2}MnO\textsubscript{3}\xspace}

\newcommand{\NTMO}{Na\textsubscript{2}Mn\textsubscript{3}O\textsubscript{7}\xspace}
\newcommand{\NMO}{NaMn\textsubscript{3}O\textsubscript{7}\xspace}
\newcommand{\LTRO}{Li\textsubscript{2}RuO\textsubscript{3}\xspace}
\newcommand{\LxRO}{Li\textsubscript{x}RuO\textsubscript{3}\xspace}
\newcommand{\LRO}{LiRuO\textsubscript{3}\xspace}
\newcommand{\LTIO}{Li\textsubscript{2}IrO\textsubscript{3}\xspace}
\newcommand{\LxIO}{Li\textsubscript{x}IrO\textsubscript{3}\xspace}
\newcommand{\LIO}{LiIrO\textsubscript{3}\xspace}
\newcommand{\LfIO}{Li\textsubscript{0.5}IrO\textsubscript{3}\xspace}

\newcommand{\LFPO}{LiFePO\textsubscript{4}\xspace}
\newcommand{\LNO}{LiNiO\textsubscript{2}\xspace}
\newcommand{\LxNO}{Li\textsubscript{1-x}NiO\textsubscript{2}\xspace}
\newcommand{\LCO}{LiCoO\textsubscript{2}\xspace}
\newcommand{\LxCO}{Li\textsubscript{1-x}CoO\textsubscript{2}\xspace}

\newcommand{\MnIV}{Mn\textsuperscript{4+}\xspace}
\newcommand{\MnV}{Mn\textsuperscript{5+}\xspace}

\newcommand{\RuV}{Ru\textsuperscript{5+}\xspace}

\newcommand{\IrVf}{Ir\textsuperscript{5.5+}\xspace}

\newcommand{\dfour}{$d$\textsuperscript{4}\xspace}
\newcommand{\dfive}{$d$\textsuperscript{5}\xspace}

\newcommand{\dfourL}{$d$\textsuperscript{4}\underline{L}\xspace}
\newcommand{\dfiveL}{$d$\textsuperscript{5}\underline{L}\xspace}
\newcommand{\dsixL}{$d$\textsuperscript{6}\underline{L}\xspace}
\newcommand{\dnplusoneL}{$d$\textsuperscript{n+1}\underline{L}\xspace}
\newcommand{\dfourLt}{$d$\textsuperscript{4}\underline{L}\textsuperscript{2}\xspace}
\newcommand{\dfiveLt}{$d$\textsuperscript{5}\underline{L}\textsuperscript{2}\xspace}
\newcommand{\dsixLt}{$d$\textsuperscript{6}\underline{L}\textsuperscript{2}\xspace}
\newcommand{\dsevenLt}{$d$\textsuperscript{7}\underline{L}\textsuperscript{2}\xspace}

\newcommand{\dfiveLthree}{$d$\textsuperscript{5}\underline{L}\textsuperscript{3}\xspace}
\newcommand{\dsixLthree}{$d$\textsuperscript{6}\underline{L}\textsuperscript{3}\xspace}
\newcommand{\dsevenLthree}{$d$\textsuperscript{7}\underline{L}\textsuperscript{3}\xspace}
\newcommand{\Ledge}{\textit{L\textsubscript{3}-}edge\xspace}
\newcommand{\Medge}{\textit{M\textsubscript{3}-}edge\xspace}
\newcommand{\Kedge}{\textit{K-}edge\xspace}
\newcommand{\ttg}{t\textsubscript{2g}\xspace}
\newcommand{\eg}{e\textsubscript{g}\xspace}

\newcommand{\figref}[2]{\hyperref[#1]{~\ref{#1}{#2}}}
\begin{document}
\title{Transition-Metal-Oxygen Covalency Controls Ligand-Hole Stability in Li-Rich Cathodes}

\author{Kuan~H.~Hsu}
\thanks{These authors contributed equally to this work.}
\affiliation{Department of Materials Science and Engineering, Stanford University, Stanford, CA, 94305, USA}
\affiliation{Stanford Institute for Materials and Energy Sciences, SLAC National Accelerator Laboratory, Menlo Park, CA, 94025, USA}

\author{Eder~G.~Lomeli}
\thanks{These authors contributed equally to this work.}
\affiliation{Department of Materials Science and Engineering, Stanford University, Stanford, CA, 94305, USA}
\affiliation{Stanford Institute for Materials and Energy Sciences, SLAC National Accelerator Laboratory, Menlo Park, CA, 94025, USA}
\affiliation{Advanced Light Source, Lawrence Berkeley National Laboratory, Berkeley, CA, 94720, USA}

\author{Jaehoon~Kim}
\affiliation{Department of Battery Engineering, Graduate Institute of Ferrous \& Eco Materials Technology, Pohang University of Science and Technology (POSTECH), Pohang, 37673, Republic of Korea}

\author{Gukhyun~Lim}
\affiliation{Department of Battery Engineering, Graduate Institute of Ferrous \& Eco Materials Technology, Pohang University of Science and Technology (POSTECH), Pohang, 37673, Republic of Korea}
\affiliation{Department of Engineering, University of Cambridge, Cambridge, CB2 1PZ, UK}

\author{Amelia Zhao}
\affiliation{Department of Materials Science and Engineering, Stanford University, Stanford, CA, 94305, USA}
\affiliation{Stanford Institute for Materials and Energy Sciences, SLAC National Accelerator Laboratory, Menlo Park, CA, 94025, USA}

\author{Wen-Shin~Lu}
\affiliation{Department of Materials Science and Engineering, Stanford University, Stanford, CA, 94305, USA}
\affiliation{Stanford Institute for Materials and Energy Sciences, SLAC National Accelerator Laboratory, Menlo Park, CA, 94025, USA}

\author{Joshua~J.~Kas}
\affiliation{Stanford Institute for Materials and Energy Sciences, SLAC National Accelerator Laboratory, Menlo Park, CA, 94025, USA}
\affiliation{Department of Physics, University of Washington, Seattle, WA, 98195, USA}

\author{John~Vinson}
\affiliation{Material Measurement Laboratory, National Institute of Standards and Technology, Gaithersburg, Maryland, 20899, USA}

\author{Brian~Moritz}
\affiliation{Stanford Institute for Materials and Energy Sciences, SLAC National Accelerator Laboratory, Menlo Park, CA, 94025, USA}

\author{Jihyun~Hong}
\affiliation{Department of Battery Engineering, Graduate Institute of Ferrous \& Eco Materials Technology, Pohang University of Science and Technology (POSTECH), Pohang, 37673, Republic of Korea}

\author{Wanli~Yang}
\email{wlyang@lbl.gov}
\affiliation{Advanced Light Source, Lawrence Berkeley National Laboratory, Berkeley, CA, 94720, USA}

\author{Thomas~P.~Devereaux}
\email{tpd@stanford.edu}
\affiliation{Department of Materials Science and Engineering, Stanford University, Stanford, CA, 94305, USA}
\affiliation{Stanford Institute for Materials and Energy Sciences, SLAC National Accelerator Laboratory, Menlo Park, CA, 94025, USA}
\affiliation{Geballe Laboratory for Advanced Materials, Stanford University, Stanford, CA, 94305, USA}

\date{\today}
\begin{abstract}

Lithium-rich cathodes boast a large specific capacity due to the presence of excess lithium per transition metal. However, making use of that excess Li drives the cathodes to unusually high formal oxidation states, often triggering unwanted side reactions, typically signaled by gas release and commonly associated with oxygen oxidation.  By varying the transition metal (TM) chemistry in Li-rich materials, cyclability can vary drastically, with some materials suffering rapid capacity fade while others are able to maintain large and stable voltage windows. Such observations indicate a crucial role for both TM and O activities, but the details remain elusive. Here, we model core-level spectroscopy to understand differences in the charge compensation mechanism between different \LMetalO (M = Mn, Ru, Ir) cathode materials. We find that ligand holes are important throughout the series, but their stability depends strongly on the transition-metal-oxygen electronic structure. In \LTMO, the ground state is already strongly negative-charge-transfer-like, so further delithiation rapidly increases ligand-hole character and promotes instability. Moving from Mn to Ru and Ir increases TM--O hybridization while reducing the tendency toward an extreme negative-charge-transfer state, allowing additional holes to remain delocalized over the covalent TM--O network over a wider range of charge. These results identify TM--O covalency, rather than the absence of oxygen-hole character, as a key factor controlling reversible high-valent redox and suggest a route for stabilizing high-energy-density cathodes without invoking reversible molecular oxygen formation.

\end{abstract}
\maketitle

\section{Introduction}

Electrochemical processes feature reduction and oxidation (redox) reactions of a variety of species across a multitude of relevant energy and industrial technologies. Transition metal (TM) oxides hold a unique place as workhorses for current and future renewable energy applications, ranging from their role as electrodes in Li-ion batteries to catalysts for hydrogen and oxygen evolution reactions. Electrochemical energy applications often trigger the formation of unconventional oxide states through a wide range of redox voltage. These states define the properties of energy materials and more importantly, their practical performance. Thus, fundamental understanding of the electronic structure and the evolution of redox active materials is paramount for the development of next generation devices.

Most challenges in developing high-performance energy storage stem from the inherent trade-off between high energy density and cycling stability in cathode materials, especially at high voltages. A representative family is lithium-rich layered oxide cathodes. Relative to conventional stoichiometric layered oxides, {\it e.g.} LiMO\textsubscript{2} (M = Co, Ni), lithium-rich cathodes feature more lithium than TM per formula unit, and therefore higher theoretical specific capacity.~\cite{goodenough_lro,Rossouw_Li2MnO3_1991,Kobayashi_Li2IrO3_2003}. However, accessing this additional capacity can lead to instabilities, which often trigger voltage hysteresis and gas release~\cite{House_trappedo2,House_o2,Bruce_O2,piper_lmo_capacity,zhou_lmo_redox,cho_lmo_overview,duda_lmo_L_rixs}. These instabilities have been attributed to the involvement of oxygen atoms in the redox reaction.

Extensive efforts have focused on stabilizing oxygen redox reactions with direct oxygen bonding to access
capacity beyond the cationic redox limit at high voltage, referred to as ``anionic/oxygen redox''~\cite{Tarascon_NE_review,Grimaud2016,Doublet_NM,bruce_non_bond_oredox}.
These efforts have faced challenges to stability and kinetics~\cite{Bruce_O2,Gent_migration,Gent_tim_review,Kim_oredox_theory,Meng_oredox_disorder}, with highly oxidized oxygen anions often forming chemically defined species, especially peroxo- and O\textsubscript{2}. Nevertheless, some studies have reported signatures of stabilized oxygen holes in $3d$ TM oxide cathodes with low hysteresis and voltage fade, even at high voltages~\cite{abate_nmo,house_superstruct,HariLFSO2025,bruce_non_bond_oredox,House_delocalO2,Li_nmo_native_vac}, although the mechanism that enables stable cycling performance remains unclear.

While achieving stable high-valent redox with $3d$ TMs is challenging, layered oxides containing heavier TMs have demonstrated the capability to undergo relatively more reversible high-valent redox processes~\cite{Sathiya_lrmo,Sathiya_lrso,Sathiya_lro_voltage_decay,kobayashi_lio_1997,doublet_lro_dft,Wu_rev_lro,Wanli_distinct_lirich,Yu_lro_xas,chan_lio_xas,mccalla_lio_dimer,hong_gent_lio,House_covalency,lro_prm_lor}. This is often attributed to charge compensation mediated by the heavy TMs. However, spectroscopic measurements on oxygen atoms reveal clear signatures of oxygen participation in the redox process with non-negligible ligand hole density~\cite{Wanli_distinct_lirich,Yu_lro_xas,mccalla_lio_dimer,House_covalency,hong_gent_lio,chan_lio_xas,lro_prm_lor}, suggesting that oxygen plays a significant role in the redox activities. In practice, the strong hybridization between the heavy TMs and oxygen makes it difficult to disentangle their individual contributions to the redox activity, resulting in a lack of consensus on the underlying mechanism that stabilizes high voltage, high-valent cycling in heavy TM oxides. Nonetheless, experimental measurements point to the same trend: heavier TM oxide cathodes can accommodate higher levels of delithiation with reversible capacity at high voltages. Thus, clarifying the TM effects on Li-rich cathode properties through a comparative study becomes crucial for understanding the fundamental mechanism that enables reversible high-voltage redox.

A natural framework for addressing this question is provided by the competition between charge-transfer energetics and TM--O hybridization. In a negative charge-transfer (NCT) regime, oxidation can be accommodated predominantly through ligand-hole configurations \dnplusoneL, rather than by further oxidation of the nominal TM ion~\cite{ZSA_diagram}. However, substantial ligand-hole character does not by itself imply a strongly negative charge-transfer energy: ligand holes can also arise from strong covalent mixing between TM $d$ and O $2p$ states. This distinction becomes increasingly important in moving from $3d$ to $4d$ and $5d$ transition metals, whose more spatially extended $d$ orbitals enhance TM--O hybridization. The stability of high-valent redox may therefore depend not simply on the number of oxygen holes, but on the electronic environment in which those holes are accommodated.

Here, we test this picture through a systematic comparison of \LMetalO (M = Mn, Ru, Ir), combining charge-transfer multiplet and first-principles calculations with experimental X-ray absorption spectroscopy (XAS) and resonant inelastic X-ray scattering (RIXS). By examining both the O \Kedge and TM \Ledge, we track the evolution of metal and ligand configurations during delithiation and distinguish ligand-hole formation associated with strong covalency from the evolution toward a strongly NCT state. The Mn-Ru-Ir series provides a particularly useful comparison because the progressively more extended TM $d$ orbitals increase TM--O hybridization while the materials exhibit markedly different ranges of reversible delithiation.

We find that ligand holes are important throughout the Mn-Ru-Ir series, but their presence alone does not determine stability. \LTMO begins in a strongly NCT regime, such that further delithiation rapidly increases ligand-hole character and drives the system toward oxygen instability. In the Ru- and Ir-based compounds, stronger TM--O hybridization produces substantial ligand-hole character even when the charge-transfer energy is less negative. The resulting holes are distributed over a more strongly covalent TM--O network, allowing progressively deeper oxidation before the system reaches the regime associated with O--O dimerization and irreversible oxygen chemistry. We therefore identify the interplay between charge-transfer energy and TM--O hybridization, rather than a simple division between “cationic” and “anionic” redox, as a key control parameter for reversible high-valent redox in Li-rich cathodes.

\section{Methods and Results}

We now examine how the balance between charge-transfer energetics and TM–O hybridization evolves across Li\(_2\)MO\(_3\) (M = Mn, Ru, Ir). We begin with \LTMO, where the localized nature of the Mn $3d$ orbitals provides the strongly NCT limit of this series, before considering the progressively more covalent Ru $4d$ and Ir $5d$ compounds.

\subsection{Manganates} \label{sec:manganates}
\begin{figure}
	\centering
	\includegraphics[width=1\textwidth]{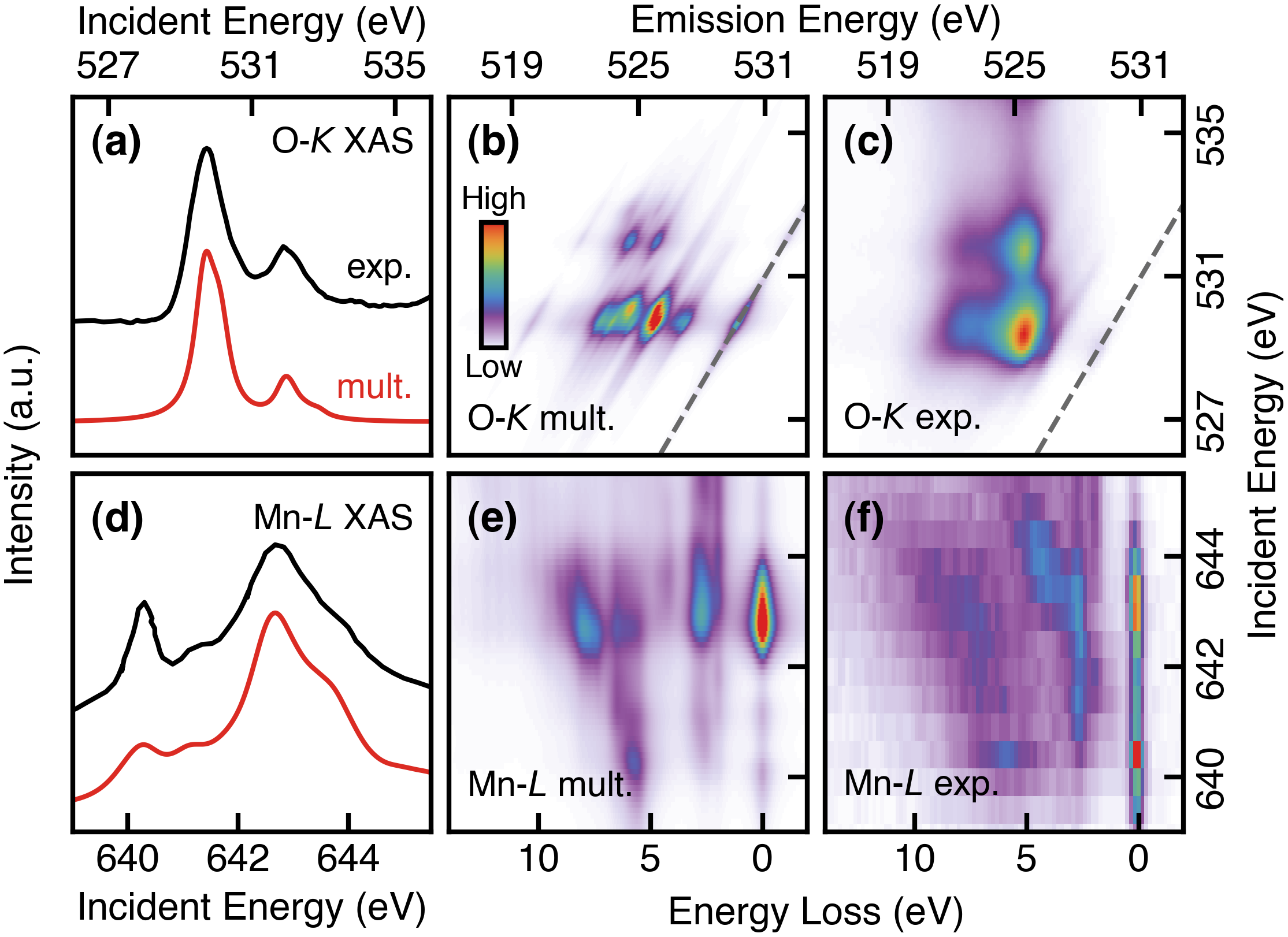}
	\caption{\textbf{Simulated and Experimental XAS and RIXS of pristine \LTMO.} (a) Comparison between experimental (total electron yield mode TEY; black) and multiplet-simulated (red) O \Kedge XAS of \LTMO. (b) Simulated and (c) experimental O \Kedge RIXS (reproduced from ref.~\cite{Wanli_distinct_lirich}) of \LTMO, plotted using incident and emission energy axes. (d) Comparison between experimental and multiplet-simulated Mn \Ledge XAS of \LTMO. (e) Simulated and (f) experimental Mn \Ledge RIXS (reproduced from ref.~\cite{duda_lmo_L_rixs}) of \LTMO, plotted as energy loss axis to highlight different $d$-$d$ and charge transfer excitations.}
	\label{fig:lmo_rixs}
\end{figure}

\LTMO provides the natural starting point for this comparison. Formally, Mn is already \MnIV in the pristine compound, leaving little capacity for further conventional Mn-centered oxidation during delithiation. Experimentally, \LTMO exhibits pronounced voltage hysteresis, structural degradation, and oxygen loss during the first charge~\cite{piper_lmo_capacity,cho_lmo_overview,zhou_lmo_redox,duda_lmo_L_rixs}. Its additional capacity has consequently been associated with oxygen-derived charge compensation, including the formation of oxygen-hole, O--O dimer, and ultimately molecular oxygen species. The key question for the present work is therefore how this oxygen activity emerges from the electronic structure of the pristine material and evolves upon further oxidation.

Here, we model pristine \LTMO using a charge transfer multiplet formalism~\cite{multiplet_code}, which allows for simulations of both the TM \textit{L\textsubscript{3}}- and O \Kedge spectra in the same cluster (see supplementary section~\ref{S-sec:multiplet}). Theoretical calculations of XAS and RIXS for both O \Kedge and Mn \Ledge show remarkable agreement with the experimental measurements (Fig.~\figref{fig:lmo_rixs}{})~\cite{zhou_lmo_redox,duda_lmo_L_rixs}. The calculation points to an NCT ground state with an even distribution of \dfourL and \dfiveLt (where \underline{L} denotes a ligand hole) with a significant number of holes on the oxygen orbitals. Signatures of the NCT ground state can be observed in the elastic scattering feature at the  O \Kedge with an incident energy around 529~eV (Fig.~\figref{fig:lmo_rixs}{(b)}). Excess hole formation within the oxygen $2p$ orbitals allows for a large coherence between the ground and intermediate states. This near elastic scattering feature at low absorption energy can also be seen in other NCT active cathode materials~\cite{lomeli2025lcolno,abate_nmo}, and stands in stark contrast to the O\textsubscript{2} RIXS feature seen around an absorption energy of 531 eV~\cite{Bruce_O2}. Additionally, fluorescent like features that can be seen around 643~eV to 645~eV in incident energy at the Mn \Ledge (Figs.~\figref{fig:lmo_rixs}{(e) and (f)}) signify the transition from a charge transfer picture with clear, localized $d$-$d$ excitations to a more strongly hybridized ground state with enhanced NCT and more itinerant electrons. First-principles calculations also show oxygen hole density present in the pristine material (Figs.~\figref{S-fig:lmo_pdos}{} and~\figref{S-fig:lmo_ocean}{}). On the other hand, clear signatures of $d$-$d$ excitations at the Mn \Ledge ($\approx$ 3~eV energy loss) point to still well localized Mn $d$ orbital states (Fig.~\figref{fig:lmo_rixs}{(e)}), which contributes to the stability of stoichiometric \LTMO. Delithiation of \LTMO leads to a further increase in the NCT character of the ground state and an over-population of ligand holes, triggering structural instability and cycling irreversibility as previously reported.~\cite{piper_lmo_capacity,cho_lmo_overview,zhou_lmo_redox,duda_lmo_L_rixs}

The prevalence of a NCT ground state with significant ligand hole density in pristine \LTMO, as seen in our calculations, is quite similar to its charged counterparts \LxNO and \LxCO, with a formal valence of TM\textsuperscript{4+} on the transition metal, except here it manifests as the starting point of the lithiated material. Delithiation leads to an increase of this NCT character. This interpretation differs from previous theoretical work, such as the proposed Mn\textsuperscript{4+}/Mn\textsuperscript{7+}~\cite{anton_mn7+} redox couple or ``non-bonding'' oxygen $p$-orbitals~\cite{cedar_lirich_nonbonding} as the main contributors to the first-charge capacity. Experimentally, oxygen outgassing and dimer formation make it challenging to model the spectroscopy and elucidate the charge compensating mechanism, leading to reports of either a lack of change in both Mn and O valence states~\cite{piper_lmo_capacity,zhou_lmo_redox} or bulk reduction due to structural degradation~\cite{duda_lmo_L_rixs}. Here, the increase of the NCT character of the ground state upon delithiation serves as a precursor for the structural instabilities and oxygen dimerization seen experimentally, in contrast to the more stable positive to negative charge transfer transition, as seen in \LCO and \LNO~\cite{lomeli2025lcolno}, or the purely positive charge transfer redox mechanism seen in \LFPO.

\begin{figure*} [!hbt]
	\centering
	\includegraphics[width=1\textwidth]{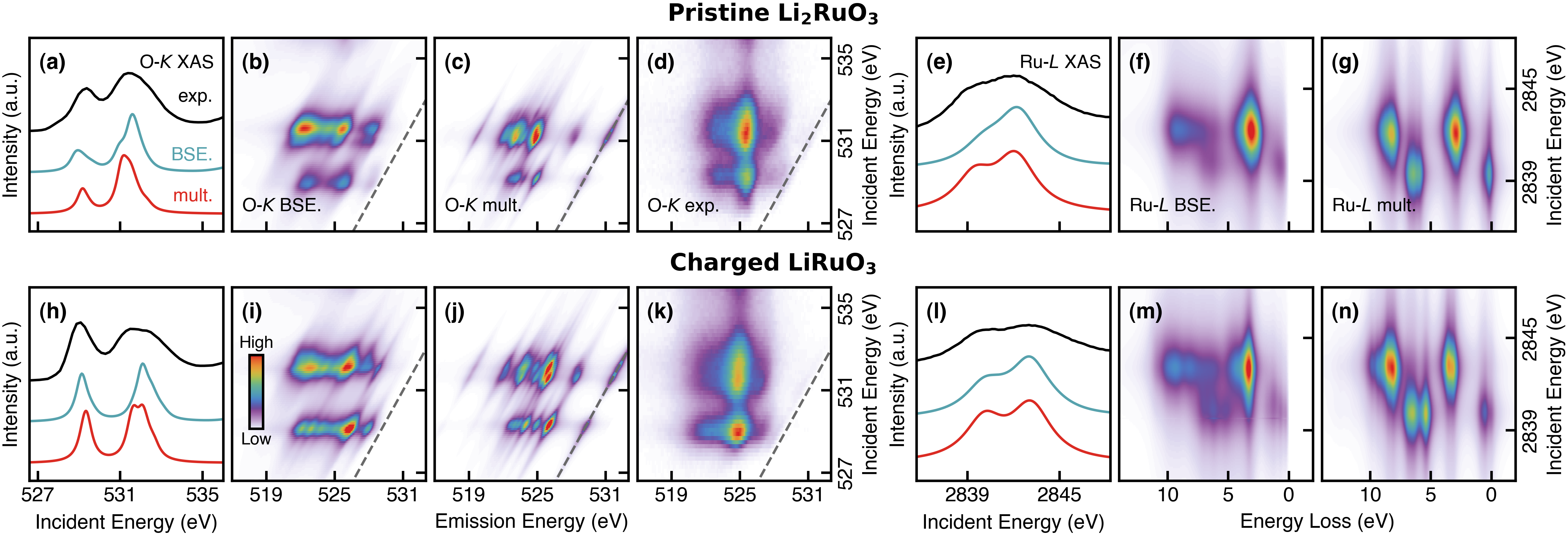}
	\caption{\textbf{Simulated and Experimental XAS and RIXS of \LxRO.} (a) Comparison between experimental (partial fluorescence yield mode PFY; black), BSE-based {\sc ocean} code simulated (blue-green), and multiplet-simulated (red) O \Kedge XAS of \LTRO. (b) {\sc ocean} simulated, (c) multiplet simulated and (d) experimental O \Kedge RIXS of \LTRO, plotted using incident and emission energy axes. (e) Comparison between experimental, {\sc ocean}, and multiplet-simulated Ru \Ledge XAS of \LTRO. (f) Simulated and (g) experimental Ru \Ledge RIXS of \LTRO, plotted using energy loss. (h-k) Experimental and Simulated O \Kedge XAS and RIXS for charged \LRO, in the same order as panels (a-d). (l-n) Experimental and simulated Ru \Ledge XAS and RIXS for charged \LRO, in the same order as panels (e-g). All experimental data in this figure are reproduced from Ref.~\cite{Wanli_distinct_lirich}. No experimental measurements of Ru \Ledge RIXS are currently available for comparison due to its tender X-ray range. \LRO was experimentally reached by cycling the electrode to 4.1 V.}
	\label{fig:lro_rixs}
\end{figure*}

We note that this evolution  of the NCT ground state in Mn cathodes can be stabilized, as seen in \NTMO,  which cycles reversibly beyond a formal valence of \MnIV~\cite{house_superstruct,abate_nmo}. For the pristine material, multiplet calculations for this system show good agreement with experimental O \Kedge RIXS and XAS (Figs.~\figref{S-fig:nmo_rixs}{(a) and (b)}), where the peak emerging at $\approx$ 527.5~eV can be associated with an increase in oxygen hole density~\cite{abate_nmo}. Meanwhile, both calculated and experimental Mn \Ledge XAS show little change in the spectral line-shape (Fig.~\figref{S-fig:nmo_rixs}{(d)})~\cite{abate_nmo}, while the calculated RIXS map shows an increase in the intensity of charge transfer excitations (Fig.~\figref{S-fig:nmo_rixs}{(e)}). For the charged nominal \MnV state, the calculated ground-state wavefunction is dominated by \dfourLt and \dfiveLthree configurations (Table~\ref{S-tab:mult_gs}). This suggests that charge compensation occurs exclusively on ligand orbitals that is driven by the increase of NCT energy. This result is consistent with previous literature, where the stability of added oxygen hole density in the material upon charging was attributed to polaronic effects~\cite{abate_nmo}.

Our calculations draw a consistent conclusion with our previous studies of conventional cathode systems ~\cite{lomeli2025lcolno}: at high formal valence, $3d$ transition metals tend to favor an NCT ground state, in which strong TM-ligand hybridization stabilizes excess ligand holes. A ground state with large NCT energy often triggers structural disorder, but can be stabilized kinetically, as in \NTMO~\cite{abate_nmo,Li_nmo_native_vac}. In both \LTMO and \NTMO, the pristine discharged cathode has the same formal valence: \MnIV. Upon charging, the voltage of \LTMO quickly increases above 4.5~V~\cite{cho_lmo_overview,piper_lmo_capacity,zhou_lmo_redox}, with observed instability and dimerization. On the other hand, \NMO operates below 4.5~V while maintaining a stable NCT ground state with significant ligand hole density~\cite{abate_nmo}. This contrast demonstrates that a ground state with large NCT energy can be accommodated without severe degradation.

\LTMO therefore represents one limit of the \LMetalO series: substantial ligand-hole character is already present before charging, and further delithiation rapidly drives the system deeper into the NCT regime. We next ask how this evolution changes when Mn $3d$ orbitals are replaced by the more spatially extended $4d$ orbitals of Ru, thereby increasing TM--O hybridization.

\subsection{Ruthenates} \label{sec:ruthenates}

Replacing Mn by the more spatially extended $4d$ orbitals of Ru provides a direct test of the role of TM-–O hybridization in stabilizing ligand-hole character. \LTRO is structurally similar to \LTMO, but exhibits substantially greater electrochemical reversibility, with low hysteresis and reversible extraction of approximately one Li per formula unit~\cite{House_covalency,Sathiya_lrmo,Sathiya_lrso,Sathiya_lro_voltage_decay,lro_prm_lor}. Further delithiation beyond the nominal \RuV state becomes increasingly irreversible~\cite{Wanli_distinct_lirich,Wu_rev_lro,lro_prm_lor}. \LTRO therefore provides an intermediate case between the rapidly destabilized Mn compound and the still more covalent Ir system considered later.

The more extended Ru $4d$ orbitals are expected to produce stronger Ru--O hybridization than in the Mn $3d$ compound. Consequently, substantial ligand-hole character does not imply that pristine \LTRO is already in a strongly NCT regime. Instead, ligand-hole configurations can acquire appreciable weight through covalent mixing between Ru $4d$ and O $2p$ states. This distinction between covalency-induced ligand-hole weight and ligand-hole self-doping associated with negative charge-transfer energy is central to understanding the different responses of \LTMO and \LTRO to delithiation.

The nature of the strong TM-ligand hybridization can be examined by core-level spectroscopy at both the O \Kedge and the Ru \Ledge, which show clear changes in spectral features upon charging~\cite{Wanli_distinct_lirich,Wu_rev_lro,Yu_lro_xas,House_covalency}. Here, we model XAS and RIXS spectra of \LxRO using both charge-transfer multiplet and the Bethe-Salpeter equation (BSE) based {\sc ocean} code~\cite{ONCVPSP,OCEAN,OCEAN3}. Due to the weaker electron-electron interactions in $4d$ orbitals~\cite{hu_ru_multiplet}, the Ru \Ledge spectra can be modeled well by both multiplet cluster and ab-initio calculations, not having the dominant multiplet splittings seen in $3d$ systems. Figure~\figref{fig:lro_rixs}{(a)} shows two absorption features around 529~eV and 531.5~eV at the O \Kedge, and broad features around 2841~eV and 2843~eV at the Ru \Ledge (Fig.~\figref{fig:lro_rixs}{(e)})). These features are associated with TM--O $\pi$ and $\sigma$ bonds (or \ttg and \eg orbitals), respectively~\cite{Wanli_distinct_lirich}, and agree well with previous experimental measurements~\cite{Wanli_distinct_lirich,Yu_lro_xas}. Theoretical O \Kedge RIXS (Figs.~\figref{fig:lro_rixs}{(b,c)}) shows good agreement compared to previous experimental results as well (Fig.~\figref{fig:lro_rixs}{(d)})~\cite{Wanli_distinct_lirich}, with a distinct two peak structure in the XAS. Additionally, calculated Ru \Ledge RIXS spectra (Figs.~\figref{fig:lro_rixs}{(f,g)}) for \LTRO reproduces the reference Ca\textsubscript{2}RuO\textsubscript{4} spectra~\cite{cro_ref_rixs}, which carries the same nominal Ru valence.

Multiplet calculations for pristine \LTRO reveal a ground-state configuration that differs qualitatively from that of \LTMO, with significant weight across \dfour, \dfiveL, and \dsixLt states (Table.~\ref{S-tab:mult_gs}). Importantly, however, this ligand-hole weight occurs even though the system is not yet strongly NCT. Instead, it arises predominantly from the large Ru $4d$-O $2p$ hybridization. First-principles calculations support this interpretation, showing substantial overlap of Ru $4d$ and O $2p$ states near the Fermi level (Fig.~\figref{S-fig:lro_pdos}). Evidence of strong hybridization in pristine \LTRO can also be seen in experimental \Ledge RIXS measurements of the reference compound Ca\textsubscript{2}RuO\textsubscript{4}~\cite{cro_ref_rixs}, displaying strong fluorescent emission signals that mirror the \LTMO \Ledge measurements in Fig.~\figref{fig:lmo_rixs}{(f)}. Thus, large ligand-hole density and a strongly negative charge-transfer energy are not equivalent: strong covalency alone can generate substantial ligand-hole character.

Upon charging to \LRO (one lithium extracted per formula unit), both the O \Kedge and the Ru \Ledge XAS exhibit similar changes in the spectra (Figs.~\figref{fig:lro_rixs}{(h) and (l)}). The separation between the two absorption features becomes larger, with the lower-energy \ttg feature increasing in intensity. While some discrepancies between the experimental and simulated spectra have been attributed to oxygen deficiency in stoichiometric \LxRO~\cite{Yu_lro_xas}, the observed spectral changes still point to electron removal from the TM-–O $\pi$-bonded orbitals, which contributes to the initial charge capacity. This is further reflected by changes in the RIXS spectra: features associated with excitation to the TM--O $\pi$ bonded orbitals, namely at an incident energy $\approx$ 528~eV at the O \Kedge (Figs.~\figref{fig:lro_rixs}{(i-k)}), and energy loss features above 5~eV at the Ru \Ledge (Figs.~\figref{fig:lro_rixs}{(m,n)}), become more pronounced with additional delithiation. We note that there are slight differences in the charge transfer excitations from the theoretical Ru \Ledge RIXS calculations, due to the localized nature of the multiplet Hamiltonian, whereas the BSE calculated RIXS shows a continuum of charge transfer excitations.  The increase in ligand hole density is reflected by a ground state reconfiguration of \LRO, where our multiplet calculations  show a more even distribution across the \dfourL, \dfiveLt, and \dsixLthree components, with slightly larger hybridization strength compared to pristine \LTRO (Table.~\ref{S-tab:mult_gs}). Similarly, figure~\figref{S-fig:lro_pdos}{(b)} shows states with strong Ru and O character, with additional oxygen 2$p$ density above the Fermi energy. A similar trend can be seen in the Ru \Medge (Figs.~\figref{S-fig:m_rixs}{(a-f)}), which shows softer X-ray transitions to the $4d$ orbitals of Ru (see supplementary section~\ref{S-sec:medge}).

The comparison with \LTMO reveals why substantial oxygen participation need not immediately produce instability. Pristine \LTRO contains appreciable ligand-hole weight, but begins from a less NCT electronic structure in which those holes are distributed over strongly hybridized Ru-O states. Delithiation can therefore be accommodated by a continuous redistribution of charge across this covalent network, involving both Ru and O character, before the system reaches the strongly NCT regime associated with oxygen instability. This behavior differs from a purely cationic-redox picture, but also from \LTMO, which begins from an already strongly NCT state.

It is important to emphasize that there is still significant ligand hole density at the onset of charging, which leads to large pre-edge features in O \Kedge XAS (Fig.~\figref{fig:lro_rixs}{(a)}). However, this is distinctly different from the cationic redox mechanism that has been suggested by previous studies~\cite{doublet_lro_dft}.  Instead, the mechanism closely resembles the NCT ground state of \LxCO and \LxNO~\cite{lomeli2025lcolno}. Here, the delocalization of the added charge allows \LRO to offset the increased NCT energy, instead of resulting in the irreversibilities associated with \LTMO. This stabilization has a finite range. Upon delithiation beyond approximately one Li per Ru, the electronic structure becomes increasingly NCT and the additional ligand-hole density is eventually accompanied by O--O dimer formation and oxygen vacancies ~\cite{Wanli_distinct_lirich,Wu_rev_lro,Yu_lro_xas,doublet_lro_dft,house_superstruct, lro_prm_lor}. Ru therefore does not eliminate oxygen instability; rather, stronger Ru-O covalency extends the range of oxidation over which ligand holes can be accommodated without irreversible oxygen chemistry. This observation suggests a natural progression to the $5d$ limit: if increasing TM--O hybridization expands this stable window, \LTIO should tolerate still deeper delithiation before reaching the corresponding instability.
\subsection{Iridates} \label{sec:iridates}

\begin{figure*} [!hbt]
	\centering
	\includegraphics[width=1\textwidth]{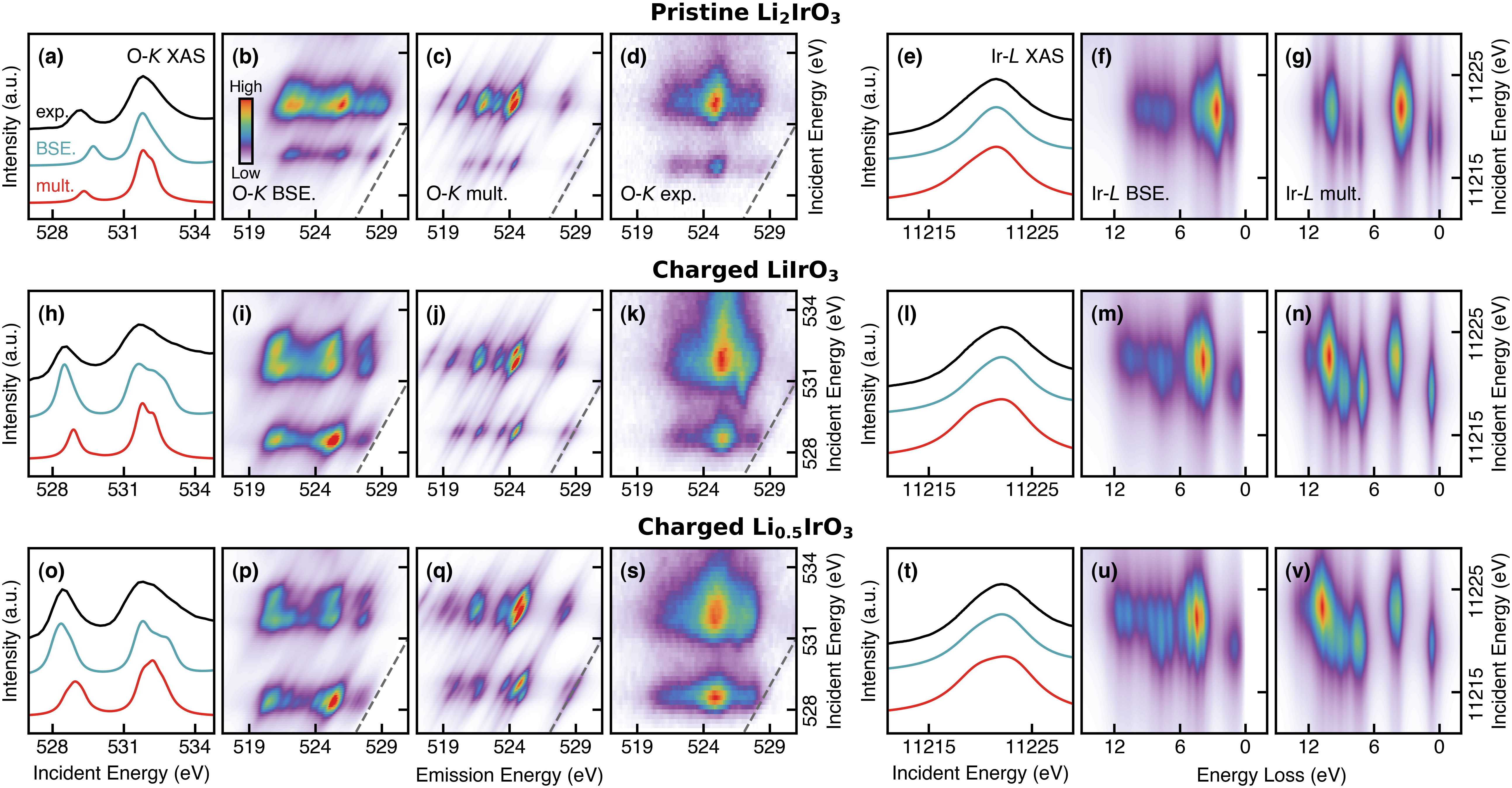}
	\caption{\textbf{Simulated and Experimental XAS and RIXS of \LTIO.} (a) Comparison between experimental (PFY; black), BSE-based {\sc ocean} code simulated (blue-green), and multiplet-simulated (red) O \Kedge XAS of \LTIO. (b) {\sc ocean} simulated, (c) multiplet simulated and (d) experimental O \Kedge RIXS of \LTIO, plotted using incident and emission energy axes. (e) Comparison between experimental, {\sc ocean}, and multiplet-simulated Ir \Ledge XAS of \LTIO. (f) Simulated and (g) experimental Ir \Ledge RIXS of \LTIO, plotted as energy loss. (h-k) Experimental and simulated O \Kedge XAS and RIXS for charged \LIO, in the same order as panels (a-d). (l-n) Experimental and simulated Ir \Ledge XAS and RIXS for charged \LIO, in the same order as panels (e-g). (o-s) Experimental and simulated O \Kedge XAS and RIXS for charged \LfIO, in the same order as panels (a-d). (t-v) Experimental and simulated Ir \Ledge XAS and RIXS for charged \LfIO, in the same order as panels (e-g). Details of experimental measurements of \LxIO at the O \Kedge are provided in supplementary section~\ref{S-sec:lio-exp}, whereas Ir \Ledge experimental data in this figure are reproduced from Ref.~\onlinecite{hong_gent_lio}. No experimental measurements are currently available for comparison with Ir \Ledge RIXS.}
	\label{fig:lio_rixs}
\end{figure*}

Extending this comparison to the $5d$ series provides a further test of the role of TM--O covalency. Among the compounds considered here, \LTIO supports the deepest reversible delithiation, with extraction of approximately 1.5 Li per Ir~\cite{kobayashi_lio_1997}. Despite reaching a formal Ir valence as high as Ir\(^{5.5+}\), the material exhibits comparatively little structural degradation and clear oxygen participation in charge compensation without formation of short O--O dimers~\cite{House_covalency,mccalla_lio_dimer,hong_gent_lio,chan_lio_xas}. Evidence for peroxide formation emerges only upon still deeper oxidation~\cite{House_covalency,hong_gent_lio}. \LTIO therefore provides an important limiting case: substantial ligand-hole character and very high formal oxidation can coexist with reversible redox over a wider range than in the Mn- or Ru-based compounds.

To evaluate this extended stability, we investigate whether the increased TM--O covalency in \LTIO allows ligand holes to be accommodated further before entering the strongly NCT regime associated with O--O bond formation. Here, we focus on the non-hysteretic region of the \LxIO ($x\geq0.5$) electrochemical cycle to test whether the trend seen in the Mn and Ru Li-rich systems holds as discussed earlier. We model O \Kedge and Ir \Ledge core-level spectroscopy using both the multiplet formalism and ab-initio calculations along with experimental measurements at the O \Kedge (see supplementary section~\ref{S-sec:lio-exp}). For pristine \LTIO, XAS spectra for both experimental and simulated O \Kedge displays the same characteristics as \LTRO, where two major features can be seen around $\approx$ 529~eV and 533~eV (Fig.~\figref{fig:lio_rixs}{(a)}) that are associated with TM--O $\pi$ and $\sigma$ bonds, respectively. On the other hand, the Ir \Ledge (and \Medge~\cite{chan_lio_xas}) only shows one broad feature (see Fig.~\figref{fig:lio_rixs}{(e)})~\cite{hong_gent_lio}, due to the short core-hole lifetime of the intermediate state; however, finer resolution may be achieved by high energy X-Ray techniques such as HERFD.

Simulated O \Kedge RIXS (Figs.~\figref{fig:lio_rixs}{(b) and (c)}) shows similar features, with excitations across the emission energy axis due to the strong hybridization (see also Fig.~\figref{S-fig:lio_pdos}{}). Multiplet calculations point to a ground state with small positive charge transfer energy, but large hybridization that contributes to the large ligand hole density with a broad distribution across the \dfive, \dsixL and \dsevenLt ground state (Table.~\ref{S-tab:mult_gs}). Indications of strong TM--O hybridization are observed in the {\sc ocean}-calculated RIXS centered at 525~eV emission energy~\cite{tom_wanli_review,chan_lio_xas,hong_gent_lio} (Fig.~\figref{fig:lio_rixs}{(b)}), with good agreement to experimental measurement~\cite{hong_gent_lio} (Fig.~\figref{fig:lio_rixs}{(d)}). Ir \Ledge RIXS (Figs.~\figref{fig:lio_rixs}{(f) and (g)}) shows broad \ttg to \eg excitations around 3~eV to 4~eV energy loss in both simulated spectra, while the multiplet calculations (Fig.~\figref{fig:lio_rixs}{(g)})  also capture $d$-$d$ excitations under 1~eV energy loss within the \ttg manifold, which qualitatively agree with prior experimental measurements~\cite{Gretarsson_ARO_rixs}. Combining O \Kedge and Ir \Ledge calculations, pristine \LTIO is therefore best viewed as a highly covalent positive-charge-transfer system rather than a strongly NCT material.

In charged \LIO, the O \Kedge low-energy peak associated with Ir \ttg orbitals grows (Fig.~\figref{fig:lio_rixs}{(h)}), while the Ir \Ledge centroid shifts to a higher incident energy (Fig.~\figref{fig:lio_rixs}{(l)}). This trend continues until charged to the formal \IrVf state (Figs.~\figref{fig:lio_rixs}{(o-v)}). The subtle spectral changes make it challenging to categorize the redox mechanism in \LxIO cathodes, with lattice oxygen redox~\cite{mccalla_lio_dimer}, TM--O hybridized redox~\cite{hong_gent_lio}, and redox on non-bonded oxygen orbitals~\cite{chan_lio_xas} all proposed as possible mechanisms. Here, theoretical O \Kedge XAS and RIXS spectra show good agreement with experimental measurements (Figs.~\figref{fig:lio_rixs}{(h-k), (o-s)}). The ground state configuration of \LIO shows a highly covalent, slightly NCT ground state with even distribution of \dfiveL, \dsixLt and \dsevenLthree (Table.~\ref{S-tab:mult_gs}). This reflects the decreased Coulomb interaction in heavy TMs when compared to a typical $3d$ TM, resulting in less negative charge transfer energy. First-principles calculations point to the same result, where the bands below the Fermi level are mostly highly covalent Ir--O states (Fig.~\figref{S-fig:lio_pdos}{}), in which oxygen contributes to the majority of charge compensation~\cite{mccalla_lio_dimer}.

While Ir \Ledge XAS shows small changes upon charging, RIXS calculations show a distinctively different loss energy profile for different \LxIO. Upon charging, the RIXS intensity shifts from \ttg to \eg de-excitation to a more charge transfer type excitation (\textit{i.e.} \dsixLt $\rightarrow$ \dfiveL), highlighting a change to the NCT ground state (Figs.~\figref{fig:lio_rixs}{(l-n)}). Further delithiation continues this trend (Figs.~\figref{fig:lio_rixs}{(t-v)}). Compared to the multiplet calculation (Figs.~\figref{fig:lio_rixs}{(g,n,v)}), {\sc ocean} simulated RIXS maps (Figs.~\figref{fig:lio_rixs}{(f,m,u)}) show a more diffusive charge transfer type excitation with loss energy from approximately 5~eV to 10~eV due to the itinerant nature of the exciton, while retaining the same trend as the multiplet calculations. Theoretical Ir \Medge RIXS shows similar changes in loss profile (Figs.~\figref{S-fig:m_rixs}{(g-o)}), and requires X-ray photon energy closer to the soft X-ray range ($\approx$ 2.5~keV), making it an ideal experiment to examine the change of ground state character in the \LxIO cathode.

Multiplet and DFT calculations for \LxIO point to a majority of charge compensation from oxygen atoms, where strong TM--O covalency and a smaller NCT energy enable stable multivalent redox up to formal \IrVf. However, as in the Ru compound, this stabilization is not unlimited. Upon oxidation beyond the formal \IrVf state, peroxide-like species have been observed~\cite{House_covalency,hong_gent_lio}, indicating that sufficiently deep delithiation ultimately drives even the highly covalent Ir-O network toward localized oxygen chemistry. Conversely, the observation that a structurally stabilized \LTIO polymorph can undergo still deeper delithiation with reduced hysteresis~\cite{pearce_beta_lio} emphasizes that the onset of instability is controlled jointly by electronic and structural energetics. Strong TM--O covalency expands the range over which ligand holes can be accommodated, while the lattice ultimately determines whether those holes remain distributed or condense into O--O species.

\section{Discussion}

The comparison of \LTMO (M = Mn, Ru, Ir) reveals that ligand-hole character is a common feature of high-valent redox across the series, but that its presence alone does not determine electrochemical stability. Instead, the three compounds differ in how ligand holes are generated and accommodated within the hybridized TM--O electronic structure. This distinction can be understood through the interplay between the charge-transfer energy, \(\Delta_{\rm CT}\), and TM--O hybridization, \(t_{pd}\). Together, these parameters determine both the ligand-hole content of the ground state and how rapidly the system evolves toward strongly negative charge-transfer character upon delithiation.

\LTMO and \LTIO illustrate the two limiting cases. In \LTMO, the ground state is already strongly NCT before delithiation, with substantial \dnplusoneL and higher ligand-hole character. Further oxidation therefore drives a rapid increase in ligand-hole density from an already NCT starting point. In \LTIO, by contrast, strong Ir--O hybridization produces substantial ligand-hole character even while \(\Delta_{\rm CT}\) remains slightly positive. Delithiation can consequently proceed through a broad range of strongly covalent states before the system becomes deeply NCT. \LTRO lies between these limits, combining stronger TM--O hybridization than the Mn compound with a less NCT pristine electronic structure. The progression from Mn to Ru to Ir therefore separates two quantities that are often conflated: ligand-hole density and negative charge-transfer character.

This distinction places the NCT ground state at the center of the redox mechanism across high-energy \LMetalO cathodes. From this perspective, different types of ``oxygen redox'', such as long peroxo bonds~\cite{mccalla_lio_dimer,Sathiya_lrso,doublet_lro_dft}, unbonded oxygen redox~\cite{cedar_lirich_nonbonding,abate_nmo,bruce_non_bond_oredox} or $\pi$-redox networks~\cite{House_delocalO2,anton_pi_redox} can all be categorized as rearrangement of ligand hole density governed by the NCT ground state. However, the NCT ground state can enable stable, non-hysteretic electrochemical cycling below 4.5~V, avoiding oxygen evolution reactions while expanding the specific energy density of a cathode material via engineered structural stability (see section~\ref{S-sec:nmo_calc} and refs.~\cite{house_superstruct,abate_nmo,HariLFSO2025}). On the other hand, oxygen dimerization and out-gassing are known phenomena when the battery voltage rises above 4.5~V~\cite{cho_lmo_overview,piper_lmo_capacity,zhou_lmo_redox,lro_prm_lor,House_covalency,lomeli2025lcolno}. While this process can be somewhat reversible~\cite{Wanli_distinct_lirich,Sathiya_lrso}, it will eventually lead to poor kinetics and capacity loss. Thus, we must prioritize stabilizing NCT ground states at high states of charge, as the large ligand hole density can easily lead to oxygen dimer formation.

Ultimately, these observations highlight the importance of the TM, as it plays a central role in dictating the stability of a high energy density cathode. At high voltages, the cathode likely exhibits an NCT ground state, where redox reactions occur on the highly hybridized TM--O. The overall stability of the cathode is dictated by the redox active TM--O framework. Cathodes that are doped with non-redox active TMs, such as Sn~\cite{House_covalency,Sathiya_lrso,hong_gent_lio,Sathiya_lro_voltage_decay}, will have a lower energy density compared to its non-doped counterpart. On the other hand, doping heavy metals into $3d$ TM based cathodes can stabilize the excess ligand hole and the NCT ground states, allowing more lithium extraction per formula unit~\cite{Sathiya_lrmo,wanli_lnmo_lnro}. Careful selection of the TM element is therefore key to designing cathode materials that combine high energy density with structural stability.

In conclusion, by using theoretical core-level spectroscopy calculations, we reveal the underlying mechanism responsible for the different electrochemical behaviors of Li-rich parent compounds with different TMs. We found the critical stabilization effect of heavy TMs is acheived through increased covalency, while NCT evolution remains the core universal redox mechanism among all materials. Here, extended $d$-orbitals lead to a less NCT ground state and stronger covalency with the ligand orbitals. This leads to greater stability at high voltages without the formation of irreversible oxidized oxygen species. This new theoretical insight provides an alternative method and opens up new design rules for stabilizing NCT states for high energy density cathodes that avoid dimerizing ``oxygen redox'' and poor electrochemical reversibility~\cite{Gent_tim_review}.

\section*{Acknowledgements}

This paper was supported in part by the U.S. Department of Energy, Office of Basic Energy Sciences, Materials Sciences and Engineering Division. Calculations were performed on the Sherlock cluster at Stanford University and on resources of the National Energy Research Scientific Computing Center (NERSC), a Department of Energy Office of Science User Facility, using NERSC award BES-ERCAP0027203. Experimental work was supported by Korea Institute for Advancement of Technology (KIAT) grant funded by the Korea Government (MOTIE) (RS-2024-00419413, HRD Program for Industrial Innovation) and the National Research Foundation of Korea (NRF) grant funded by the Korea government (MSIT) (RS-2025-00517846). Spectroscopic experiments are performed at Beamline 8.0.1.1 (iRIXS) and 6.0.2 (QERLIN) of the Advanced Light Source, a U.S. DOE Office of Science User Facility under contract no. DE-AC02-05CH11231.

Certain software is identified in this paper in order to specify the experimental procedure adequately. Such identification is not intended to imply recommendation or endorsement of any product or service by NIST, nor is it intended to imply that the materials or equipment identified are necessarily the best available for the purpose.

We thank William Chueh, Will Gent and Edward Mu for helpful discussions.

\subsection{Author Contributions}
T.P.D and W.Y. conceived the study. K.H.H and E.G.L. performed the DFT calculations and data analysis under the supervision of B.M., and T.P.D.. K.H.H and E.G.L. performed the spectroscopic XAS and RIXS simulations under the supervision of J.V., J.J.K., B.M., and T.P.D.. J.K, G. L., J. H. and W.Y. conducted synchrotron experiments and sample preparation. A.Z. and W.-S.L. conducted data analysis and revisions for the manuscript. K.H.H and E.G.L. wrote the first draft of the manuscript and all authors reviewed and approved the final version.

\subsection{Competing Interests}
The authors declare no competing interests.

\subsection{Data Availability}
Data is available from the authors upon reasonable request.

\bibliography{main}

\end{document}